\documentclass[prl,floatfix,preprint]{revtex4}

\usepackage{graphicx}
\usepackage{amssymb}

\newcommand{\vs}{\mathbf{s}}

\newcommand{\degr}{^\circ}

\newcommand{\gBS}{\gamma_{BS}}
\newcommand{\gLIA}{\gamma_{LIA}}
\newcommand{\gC}{\gamma_c}

\begin{document}

\title{Cascade of vortex loops initiated by a single reconnection of quantum
vortices}
\author{Miron Kursa}
\affiliation{Interdisciplinary Centre for Mathematical and Computational Modelling,
University of Warsaw, Poland}
%\email{M.Kursa@icm.edu.pl}
\author{Konrad Bajer}
\affiliation{Institute of Geophysics, Faculty of Physics, University of Warsaw, Poland}
%\email{kbajer@fuw.edu.pl}
\author{Tomasz Lipniacki}
\affiliation{Institute of Fundamental Technological Research, Warsaw, Poland}
\email{tlipnia@ippt.gov.pl}

\begin{abstract}
We demonstrate that a single reconnection of two quantum vortices can lead to creation of a cascade of vortex rings. 
Our analysis, motivated by the analytical solution in LIA, involves high-resolution Biot-Savart and Gross-Pitaevskii simulations.
The latter showed that the rings cascade starts on the atomic scale, with rings diameters orders of magnitude smaller than the characteristic line spacing in the tangle. 
So created vortex rings may penetrate the tangle and annihilate on the boundaries. 
This provides an efficient mechanism of the vortex tangle decay in very low temperatures.
\end{abstract}

\maketitle
\begin{figure}[t]
\centering
\includegraphics[height=0.3\textwidth]{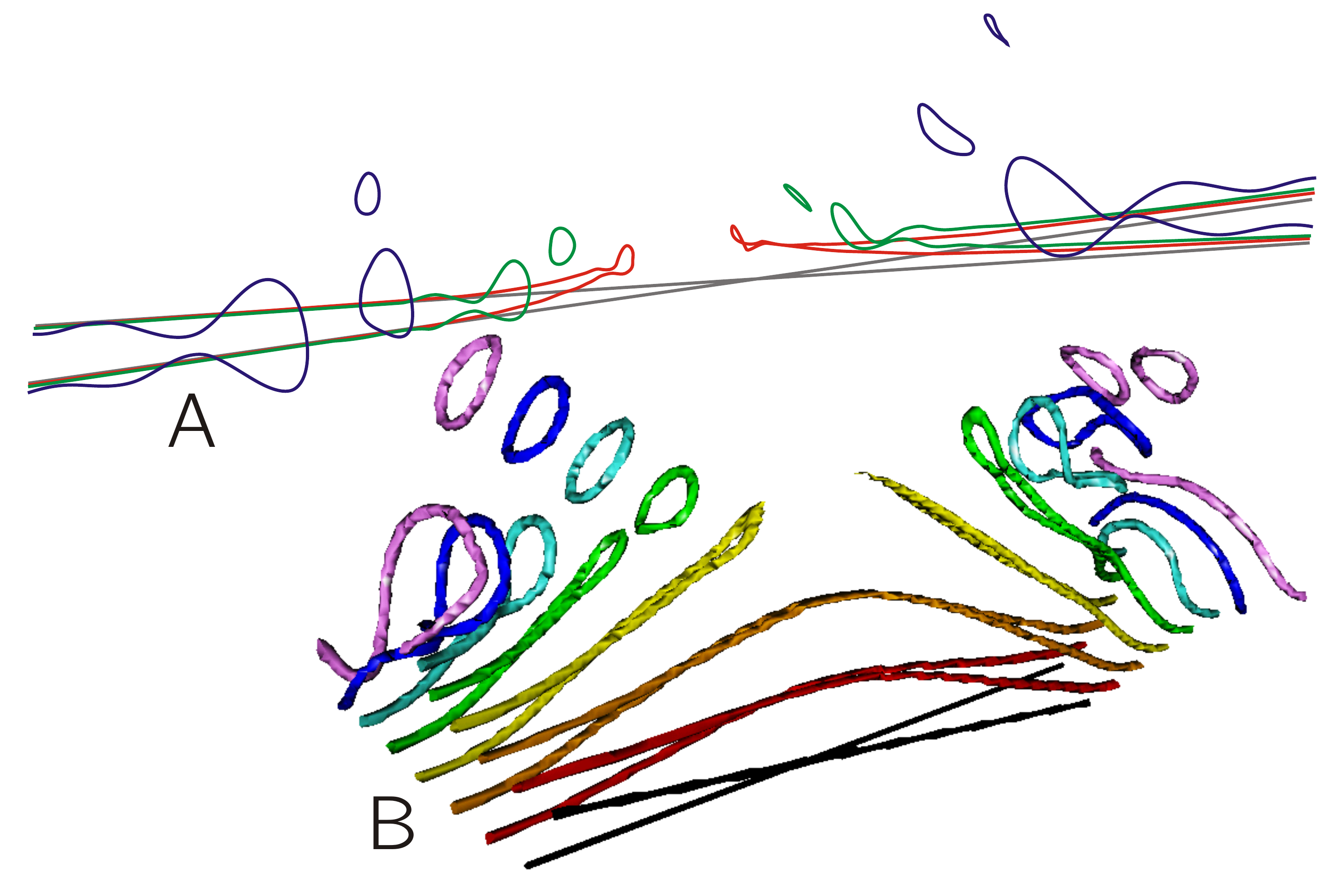} 
\caption{
Biot-Savart A) and Gross-Pitaevskii B) simulations of the reconnection of two initially straight vortex filaments in pure superfluid ($\alpha=0$), inclined at the angle $\gamma=5\degr$. 
In the BS simulation the initial separation is of $2 \times 10^{3} a_s$; the sharp corner arising at the reconnection is smoothed by the arc of radius equal to $20 a_s$.
In the GP simulation the isosurfaces of $|\Phi |^2=0.3$ are shown.
The initial line separation is equal $4 a_s$; the two arising rings have radii of the order of $3 a_s$.
}
\label{fig:3}
%\vspace{-0.5cm}
\end{figure}

We consider quantum vortex lines of constant circulation $\kappa$ (for the superfluid $^4$He $\kappa=h/m_{He}=9.97\times 10^{-4}\mathrm{cm}^2/\mathrm{s}$).
The curve traced out by a vortex filament is specified in the parametric form $\vs(\xi,t)$, with $t$ and $\xi $ denoting respectively time and arc length.
The vortex local velocity $\dot{\vs}(\xi,t)$, given by the Biot-Savart (BS) integral, in some cases can be approximated in terms of the, so called, localized induction approximation (LIA) retaining only the effects of the local vortex curvature \cite{Schwarz1985}; which in nondimensional units reads 
\begin{equation}
\dot{\vs}(\xi,t)=\vs'\times \vs''+\alpha \vs'',  \label{5}
\end{equation}
where the overdot and the prime denote the derivatives with respect to $t$ and $\xi$ respectively, and $\alpha$ is the non-dimensional friction parameter.
For $\alpha>0$, as showed by Lipniacki \cite{Lipniacki2000,Lipniacki2003}, Eq. \ref{5} has four-dimensional class of self-similar solutions.
When the initial vortex configuration consists of two half-lines with a common origin, the line motion is equivalent to a homothety transformation $\vs(\xi,t)$ $=\vs(l)\sqrt{t},$ with $l:=\xi /\sqrt{t}$. 
In terms of curvature $c(\xi,t)=K(l)$ and torsion $\tau (\xi,t)=T(l)$ the self similar solution can be given in the implicit form 
\begin{equation}
l(K)=\pm 2\sqrt{\frac{\alpha ^2+1}{\alpha}}\int_{K}^{K_{0}}\frac{dk}{k\sqrt{\ln (K_{0}/k)+\alpha (K_{0}^2-k^2)}},  \label{6}
\end{equation}
\begin{equation}
T=-\frac{K'}{\alpha K},  \label{7}
\end{equation}
where $K_{0}=K(l=0)$.
In the limit of $\alpha \rightarrow 0$ this solution converges to the self-similar solution found by Buttke \cite{Buttke1988}, and analyzed by Svistunov \cite{Svistunov1995} 
\begin{equation}
c=\frac{c_{0}}{\sqrt{t}},\quad \tau=\frac{\xi}{2t}.
\end{equation}
%%%%%%%%%%%%%%%FIG1%%%%%%%%%%%%%%%%%%%%%%%%%%
\begin{figure}[t]
\centering
\includegraphics[width=0.4\textwidth]{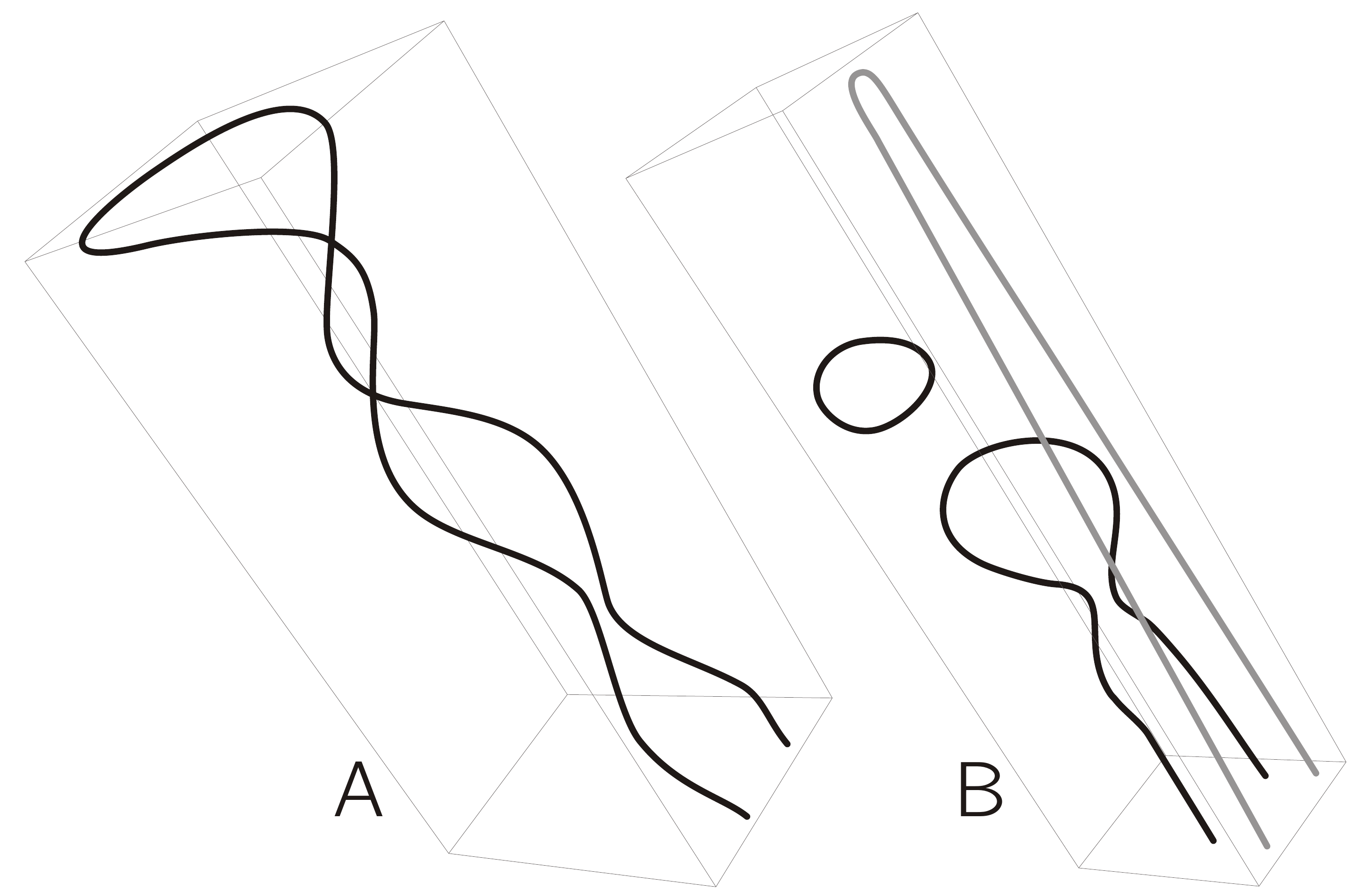} 
\caption{ 
A) Self-similar solution in the LIA approximation of one of the two vortices created by the reconnection of two straight vortex lines at the angle $\gamma=5^\circ$ for $\alpha =0.01$.
The presented `unphysical' solution has two self-crossings. 
B) Generation of a vortex ring in a Biot-Savart law simulation due to the self-reconnection of the vortex filament created by the reconnection of two straight vortex lines for the same $\gamma$ and $\alpha$ as in A). 
The initial configuration is shown as a grey line.
}
\label{fig:1}
\end{figure}
%%%%%%%%%%%%%%%%%%%%%%%%%%%%%%%%%%%%%%%%%
We reconstructed vortex lines given by the solution (\ref{6}-\ref{7}) using the Frenet-Seret equations to show that for sufficiently small $\alpha$ ($\alpha<0.44$) and sufficiently small angle between the reconnecting lines $\gamma<\gLIA(\alpha)$ the resulting vortex line has two or more self-crossings, see Fig.~\ref{fig:1}A. 
Such solutions are not physical, but their existence suggests that the reconnection of two straight vortex lines at a sufficiently small angle may lead to a series of vortex self-reconnections and the creation of a cascade of vortex rings of growing diameter, as predicted at $T=0$ by Svistunov \cite{Svistunov1995}. 
We confirmed creation of vortex rings cascades by performing high-resolution BS numerical simulations (following the numerical method proposed by Aarts \cite{Aarts1993}), starting from the configuration which arises shortly after the reconnection of two straight lines Fig.~\ref{fig:1}B.
To avoid singularity, the sharp corner in the initial configuration was replaced by an arc of radius three time larger than the radius of the vortex core, $a_{s}\approx 1.3$\AA. 
In the example simulation showed in Fig.~\ref{fig:1}B, performed for $\alpha=0.1$ and $\gamma=5\degr$, we observe the creation of a vortex ring. 
%%%%%%%%%%%%%%%FIG2%%%%%%%%%%%%%%%%%%%%%%%%%%
\begin{figure}[t]
\centering
\includegraphics[height=0.3\textwidth]{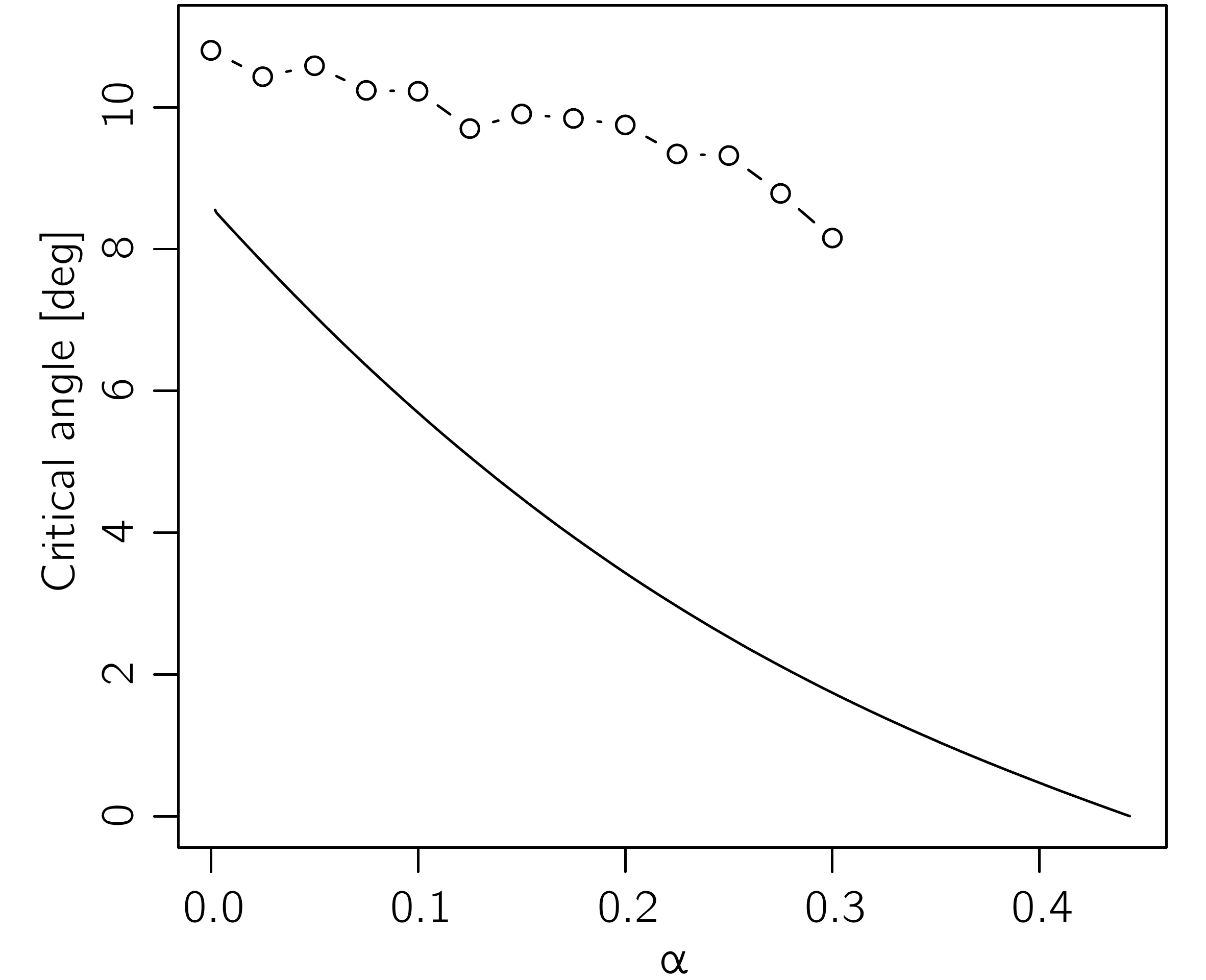}
\caption{ 
The critical angle $\gLIA(\alpha)$ below which the analytical solutions have at least two self crossings and the computed critical angle $\gBS(\alpha)$ below which vortex rings are generated in the Biot-Savart simulations. 
For $\alpha \gtrsim 0.3$ nonlocal interactions result in the collapse of a whole vortex rather than in the creation of vortex rings.
}
%\vspace{-0.5cm}
\label{fig:2}
\end{figure}
%%%%%%%%%%%%%%%%%%%%%%%%%%%%%%%%%%%%%%%%%
In Fig.~\ref{fig:2} we compare the critical angle $\gLIA(\alpha)$ below which analytical solutions have at least two self crossings and the estimated critical angle $\gBS(\alpha)$ below which the vortex rings are generated in the BS simulations. 
Non-local interactions present in the BS simulations enable ring generation for even larger $\gamma$ than expected from the LIA solutions ($\gBS(0)\approx 10.4\degr$ versus $\gLIA(0)\approx 8.5\degr$). 
In addition the non-local interactions cause two straight nearly antiparallel vortex lines to approach closely which allow them to undergo reconnection and initiate the vortex loops cascade, Fig.~\ref{fig:3}A.

Since the BS dynamics may describe the vortex motion before and after the reconnection but not the reconnection event itself, we repeated the simulation of the reconnection of two nearly antiparallel vortices based on the Gross-Pitaevskii (GP) equation, implementing the Dufort-Frankel scheme described in  \cite{Lai2004}. 
As shown in Fig.~\ref{fig:3}B the first ring arising in the cascade has the atomic scale radius of the order of $3 a_{s}$.  
This explains why the phenomenon was overlooked in the large scale BS simulations of the tangle in which the cusps arising after reconnections are replaced by the arcs of radii comparable to the characteristic radius of curvature of lines in the tangle. Interestingly, in one of the first reconnection studies based on the GP equation, Koplik and Levine \cite{Koplik1993} showed a transient configuration, possibly preceding the separation of a tiny vortex loop, which either quickly collapsed or left the small simulation box.

\begin{figure*}[tb]
\centering
\par
\includegraphics[width=1\textwidth]{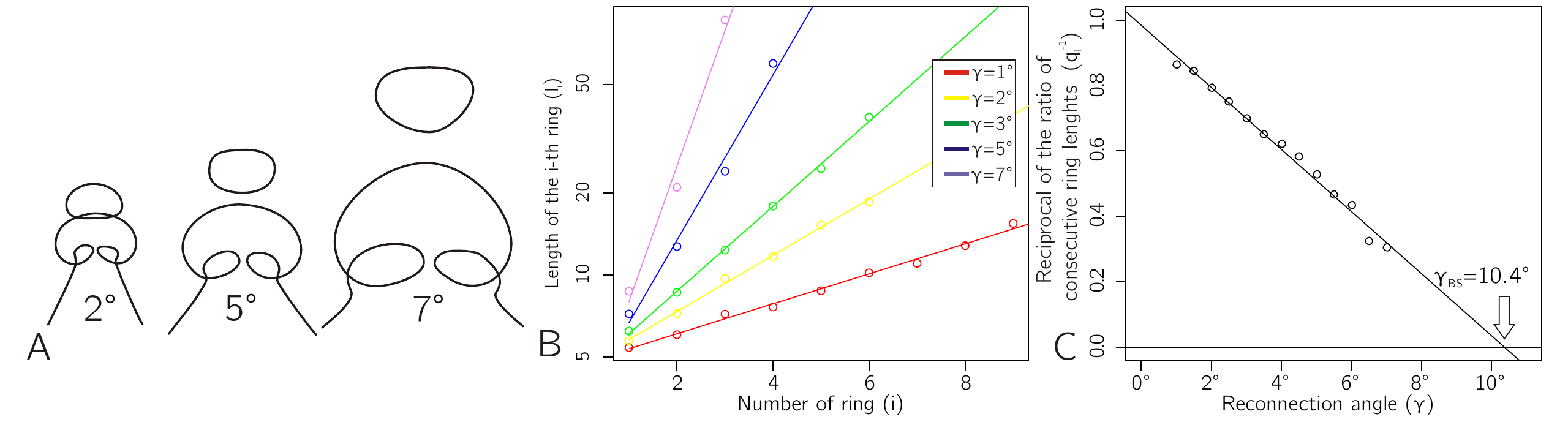} 
\caption{Rings generation after the reconnection of two straight vortex lines (BS simulations at $T=0$).
A) vortex configuration prior to the second ring separation for reconnection angles $2\degr$, $5\degr$, $7\degr$. 
B) The lengths $l_i$ of the subsequent rings for the reconnection angles $\gamma=1\degr$, $2\degr$, $3\degr$, $5\degr$, $7\degr$ are shown in logarithmic scale.
C) Inverse ratio of lengths of the subsequent rings  ($1/q_l$, cf. Eq. \ref{10}) as a function $\gamma$. The critical reconnection angle $\gBS$ is marked.
}
\label{fig:4}
\end{figure*}

\begin{figure}[t]
\centering
\includegraphics[width=0.4\textwidth]{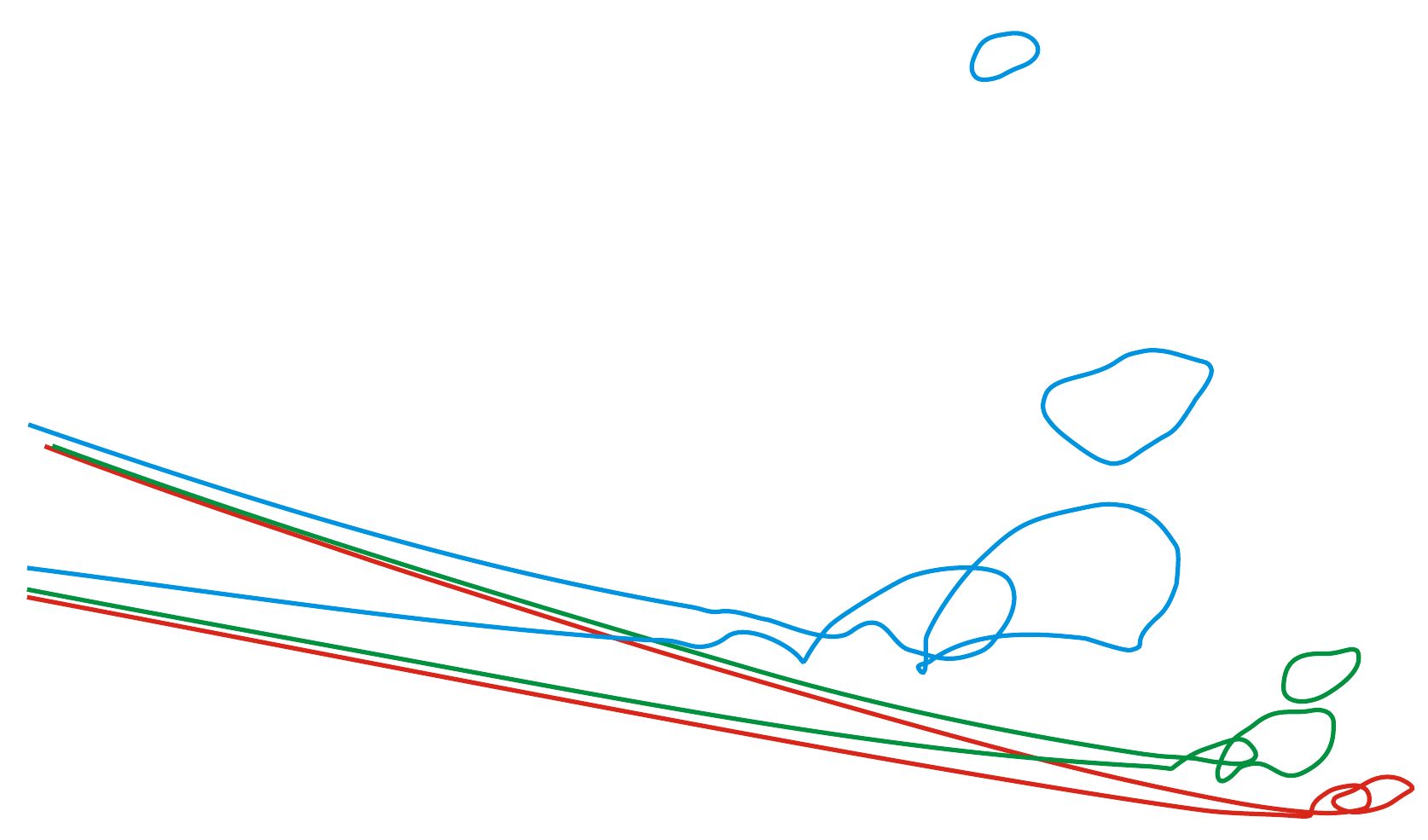}
\caption{
The BS simulations at $T=0$ showing rings generation after the reconnection of two co-planar rings with radii of $10^3 a_s$ at angle $\gamma=5\degr$. 
The rings cascade progresses until the angle between the two arms exceeds the critical value $\gBS$; three out of four created rings are shown. 
}
\label{fig:5}
\end{figure}

In further analysis we focus on the zero-temperature limit ($\alpha=0$) and restrict ourselves to numerically more efficient BS simulations. 
It follows from the similarity of the subsequent vortex rings, and is confirmed by the numerical simulations (see Fig.~\ref{fig:4}B), that their lengths $l_i$ form a geometric sequence, i.e. 
\begin{equation}
\frac{l_{i+1}}{l_i}=q_l(\gamma).  \label{10}
\end{equation}
The simulations shown in Fig.~\ref{fig:4}A indicate that $q_l(\gamma)$ is a monotonically growing function of the reconnection angle $\gamma$ diverging as $\gamma\rightarrow\gBS$ and may be well approximated as $q_l(\gamma)=\gBS/(\gBS-\gamma)$, see Fig.~\ref{fig:4}C. 
Correspondingly, the times of the subsequent ring separations also form a geometric sequence
\begin{equation}
\frac{t_{i+1}}{t_i}=q_{t}(\gamma),\quad \mathrm{where}\quad q_{t}(\gamma)\cong q_l^2(\gamma).
\end{equation}
The last equality, implies that the line-length $l(t)$ which evaporates in the form of vortex rings during time $t$ after the reconnection of two straight vortex lines is (in dimensional units) $l(t)\cong c(\gamma)\sqrt{\kappa t}$, with numerically estimated $c(\gamma)=2.5\pm 0.3$ for $0<\gamma<\gBS$.
%%%%%%%%%%%%%%%%%%%%FIG3%%%%%%%%%%%
%%%%%%%%%%%%%%%%%%%%%%%%%%%%%%%%%%%
Having numerically determined $q_l(\gamma)$ one can calculate the line-length which evaporates in the form of vortex rings $\sum_il_i\approx q_ll_{n}/(q_l-1)$, where $l_{n}$ is the length of the last ring of the cascade. 
For an idealized reconnection of two straight vortex filaments in infinite volume the cascade would be infinite. However, in the realistic situation of the reconnections of curved vortex lines in the tangle, the cascade of rings will be terminated when the angle between the two lines becomes larger than the critical angle $\gBS$. 
In Fig ~\ref{fig:5} we show the most unfavorable configuration in which the angle between the reconnecting lines grows fastest. 
Assuming that vortex lines with radius of curvature $r$ reconnect at angle $\gBS/2$, we may give the lower bound (based on the most unfavorable configuration) of the evaporated line length as $l_{tot}=2\pi r\gBS/360\degr$. 
The characteristic radius of curvature for the line in the tangle is $ r=c_{1}^{-1}L^{-1/2}$, where $c_{1}$ is the nondimensional coefficient of curvature, introduced and estimated by Schwarz \cite{Schwarz1988}.
%%%%%%%%%%%%%%%%%FIG5%%%%%%%%%%%%%
%%%%%%%%%%%%%%%%%%%%%%%%%%%%%%%%%%
%%%%%%%%%FIG4%%%%%%%%%%%
%%%%%%%%%%%%%%%%%%%%%%%%%%%%%%

To estimate the effective vortex tangle decay rate in the $T\rightarrow 0$ limit we distinguish two regimes of 1) `optically' transparent and of 2) `optically' opaque tangle. 
For the first regime we assume that all created rings are annihilated at the boundaries, and as a result the average line length loss due to each reconnection is $l_{tot}=2\pi \gBS c_{1}^{-1}L^{-1/2}/360\degr \simeq 0.09c_{1}^{-1}L^{-1/2}$. 
For the optically opaque tangle only the smaller rings may reach the boundary and annihilate. 
The mean free path $S$ of a ring of diameter $d$ in the tangle of density $L$ can be estimated as $S=1/(dL)$. 
Assuming, for simplicity, that only rings having the mean free path longer than half of the vessel diameter $D/2$ (i.e. rings having diameter $d<2/(DL)$) will reach the boundary, we calculate the maximum possible line-length loss associated with a single reconnection by taking $l_{n}=2\pi /(DL)$ 
\begin{equation}
l_{tot}^{\ast}(\gamma):=\frac{2\pi}{DL}\frac{q_l(\gamma)}{q_l(\gamma)-1}.  \label{11.5}
\end{equation}
Now, if $l_{tot}^{\ast}(\gBS)>l_{tot}$ (which implies $\gBS c_{1}^{-1}L^{1/2}D<360\degr $) the tangle is optically transparent. 
Otherwise the tangle is optically opaque. 
The reconnection frequency per unit volume  is $f_{r}=c_{1}\beta L^{5/2}$ \cite{Barenghi2004}. 
Assuming the random distribution of the reconnection angles, the fraction of reconnections at which vortex cascades are created is $\mu=\gBS/180\degr$ and thus we obtain the decay rate of an optically transparent tangle as 
\begin{equation}
\frac{dL}{dt}=-\mu f_{r}l_{tot}\simeq -0.01\beta L^2,  \label{12}
\end{equation}
and for an optically opaque tangle ($\gBS c_{1}^{-1}L^{1/2}D<360\degr$) as
\begin{eqnarray}
\frac{dL}{dt} &=&-\mu f_{r}\frac{1}{\gBS}\left( \int_{0}^{\gC}l_{tot}d\gamma +\int_{\gC}^{\gBS}l_{tot}^{\ast}(\gamma)d\gamma \right)   \nonumber \\ &=&-\frac{2\pi \mu \beta c_{1}L^{3/2}}{D}\left(1+\ln\left( \frac{\gBS L^{1/2}D}{360\degr c_{1}}\right) \right),  \label{13}
\end{eqnarray}
where $\gC=\frac{360\degr c_{1}}{DL^{1/2}}$.
As found already by Schwarz \cite{Schwarz1985}, even if the lines are initially inclined at larger angle they reorientate so that at the moment of reconnection they become almost antiparallel. 
As a result one may expect that $\mu $ is substantially larger than $\gBS/180\degr$ and thus the decay rate in Eq. \ref{12} should be regarded as a lower bound. 
The vortex tangle decay rate for the optically transparent tangle regime has the same form as the decay term in the classical Vinen equation, which in Schwarz's notation \cite{Schwarz1988} reads $[dL/dt]_{dec}=-\alpha c_{2}^2\beta L^2$, where $c_{2}^2$ is the coefficient of averaged squared curvature. 
According to Schwarz's simulations \cite{Schwarz1988} for a steady state tangle $\alpha c_{2}^2=0.12$, for the smallest studied $\alpha=0.01$  (corresponding to $T=1.07K$). 
Since at low temperatures, $\alpha\sim T^{5}$ we may expect that dissipation due to rings generation exceeds the friction force dissipation at temperature below $T\simeq 0.65K$.
%\section{Conclusions}
Buttke's solution \cite{Buttke1988}, implies that the wave number $k$ of the Kelvin helical waves at $\xi=\sqrt{\kappa t}$ decreases as $1/\sqrt{\kappa t} $, while the amplitude $A$ increases as $\sqrt{\kappa t}$. Since the energy per unit length radiated by sound (for dipole radiation, neglecting logarithmic terms, see \cite{Vinen2000}) is $\Pi=\rho \kappa ^{5}k^{6}A^2/(64\pi c_{s}^2)$ (where $c_{s}$ is sound velocity) the rate of energy dissipation and line length reduction decreases as $t^{-2}$. Consequently the line length is lost due to sound emission immediately each after reconnection event. The associated decay rate is $dL/dt=-f_{r}l_{s}=$ $-c_{1}\beta l_{s}L^{5/2}$, where $l_s$ is line loss associates with single reconnection. Assuming that $l_{s}$ is of order of $\kappa /c_{s}$ (which defines the spatial scale of the effective sound emission) we obtain that the Kelvin wave dissipation mechanism dominates over the dissipation via rings cascades generation when
\begin{equation}
pc_{1}\frac{\kappa L^{1/2}}{c_{s}}>0.01,\label{14}
\end{equation}
where $p$ is a constant.
The analogous estimation was derived by Vinen \cite{Vinen2000} who found that Kelvin wave dissipation dominates over the mutual friction dissipation (which has the same form as the dissipation rings cascades) for  
\begin{equation}
\frac{\kappa L^{1/2}}{c_{s}}>\alpha,
\label{15}
\end{equation}
which for $L=10^{6}/cm^2$ gives the critical temperature of $T\simeq 0.65K$. 
Consequently, inequalities \ref{14}-\ref{15} imply that there are three regimes: i) high-temperature regime ($\alpha >\kappa L^{1/2}/c_{s}$) when mutual friction dominates; ii) the low-temperature, dense tangle regime ($\alpha <\kappa L^{1/2}/c_{s}$, $L>0.01c_{s}/pc_{1}\kappa \simeq 10^{6}$) when  Kelvin wave dissipation dominates and iii) the low-temperature, sparse tangle regime ($\alpha <\kappa L^{1/2}/c_{s}$, $L\lesssim 10^{6}$) when the most efficient mechanism of dissipation is the generation of vortex rings. 

The question whether $T\rightarrow 0$ is a singular or regular limit of quantum turbulence is still not resolved.
Experiments on quantum turbulence imply that some dissipative mechanisms persist even at the lowest attainable temperatures \cite{Vinen2007}. 
The mechanisms invoked to explain this dissipation in pure superfluid are all associated with vortex reconnection. 
Reconnections cause the direct line loss \cite{Leadbeater2001} and they trigger Kelvin waves propagating along the reconnected vortices  \cite{Kozik2009,Svistunov1995,Leadbeater2003}. 
The nonlinear cascade of such waves may possibly transfer energy to the smallest scales where it is dissipated by the emission of phonons \cite{Leadbeater2003,Barenghi2008}. 
Finally, as demonstrated here, on the basis of the fine-scale Biot-Savart and Gross-Pitaevskii simulations, vortex reconnections at sufficiently small angle ($\gamma<10.4\degr$) lead to the creation of cascades of vortex rings of diameters starting from the atomic scale. 
Creation of rings cascades introduces the qualitative difference in the line length dissipation. 
Although, the distribution of wave numbers corresponding to subsequent rings and that of Kelvin waves arising at larger reconnection angles are similar, the  time evolution of Kelvin waves and of rings are different; the wave vectors of Kelvin waves quickly decrease in time and thus the sound emission stops, while the curvatures of vortex rings grow in the course of energy dissipation thus increasing the dissipation rate. 

In conclusion, creation of vortex rings cascade provides a novel, efficient mechanism of line-length loss at very low temperatures which is not based on acoustic dissipation. Ring creation and emission is an old idea. However, starting from the hypothetical Feynman's scenario in which vortex rings were assumed to split into smaller rings it was always assumed that energy cascades from small to large wavelengths. 
The evaporation via vortex ring generation has been considered by Barenghi and Samuels \cite{Barenghi2002}. 
However, also in this mechanism the radii of the generated vortex rings are of the order of the characteristic vortex line curvature in the tangle. Such vortex rings may escape only from a small, localized packet of vorticity.
Here we showed that the classical large-to-small-scale cascade can be bypassed when the reconnecting vortex lines are at least locally nearly antiparallel. 
A sequence of vortex rings is then produced with the diameters starting at the scale of the core size and increasing up to the scale of characteristic radius of curvature in the tangle. 
Opposite to the normally envisaged cascade (sometimes called Richardson cascade) this is a mechanism of direct line-length transfer to the smallest spatial scales.

\bibliographystyle{apsrev}

\end{document}